\documentclass{aa}
\usepackage{graphics}  

\newcommand{\gx}{{\rm GX}\,339$-$4}
\begin{document}


   \thesaurus{06     
              (08.02.1;  
                              08.09.2)  
                       } 
   \title{MOST Radio Monitoring of GX 339$-$4}


   \author{D.\ C.\ Hannikainen,
          \inst{1}
           R.\ W.\ Hunstead\inst{2},  D.\ Campbell-Wilson\inst{2}
          \and
           R.\ K.\ Sood\inst{3}}

   \offprints{D.C. Hannikainen}

   \institute{Observatory, PO Box 14,
              00014 University of Helsinki, Finland \\
              email: diana@carina.astro.helsinki.fi
        \and
	      School of Physics, University of Sydney, NSW 2006, Australia 
        \and
              School of Physics, Australian Defence Force Academy, Canberra, ACT 2600, Australia 
              }

   \date{Received 27 April 1998; accepted }

   \titlerunning{Radio Monitoring of GX 339-4}

   \authorrunning{D. Hannikainen et al.}

   \maketitle

   \begin{abstract}

The Molonglo Observatory Synthesis Telescope (MOST) has been
monitoring the candidate Galactic black hole binary system
\object{{\rm GX}\,339$-$4} at 843 MHz since 1994 April.  We present
the results of this program up to 1997 February and show a
possible correlation between radio and X-ray light curves.

   \end{abstract}

%

\section{Introduction}
The low mass X-ray binary {\gx} was discovered by the OSO-7 satellite
in 1973 (\cite{mar73}). It has been classified as a black hole
candidate primarily because of the similarity of its X-ray emission to
that of the canonical black hole system \object{{\rm Cyg}\,X-1}
(bimodal X-ray states: high/soft and low/hard) and because of rapid
variability in its X-ray and optical emission (e.g., \cite{mak86};
\cite{miy92}; \cite{now95}).

{\gx} exhibits four distinct X-ray states, three of which were
initially identified by Markert et al.\ (1973): high, low and off.
The high state is characterized by an extremely soft spectrum ($kT$ =
1--2 keV) accompanied by a hard power-law tail, the low state is
described by a single power-law hard spectrum, and the off state is in
fact a very weak hard state (\cite{mot85}).  Recently an intermediate
state between the low and the high states has been reported
(\cite{men97}).

The optical counterpart was identified as an 18 mag star
(\cite{dox79}; \cite{cow91}) which was found to be highly variable,
with $V$ ranging from $\sim$15.4 to $>$20. Photometric data revealed a
14.8 hour modulation which has been attributed to the orbital period
(\cite{cal92}). Emission from the accretion disk has dominated the
spectra making it difficult to obtain a definitive estimate for the
mass of the compact object (\cite{cow87}).  Distance estimates vary
from 1.3 kpc (\cite{pre91}) to $\sim$4 kpc (\cite{mak86}).

Simultaneous optical and X-ray observations have
shown quasi-periodic oscillations (QPOs) at mean periods of $\sim$10\,s
and $\sim$20\,s (\cite{mot83}) in the X-ray low state (see
\cite{t&l95} and references therein).  The relationship between
optical and X-ray fluxes is not well understood. For example, in 1981 Motch et
al.\ (1983) found anticorrelation between the 1--13 keV X-ray and
optical fluxes, but correlation between the 13--20 keV X-ray and
optical.

The discovery of a variable radio counterpart (\cite{s&cw94})
initiated the monitoring program at 843 MHz undertaken with the
Molonglo Observatory Synthesis Telescope (MOST) which is the subject
of this paper.  Fender et al.\ (1997) observed {\gx} at high
resolution in 1996 July with the Australia Telescope Compact Array
(ATCA) at a wavelength of 3.5 cm and reported the detection of a
jet-like extension to the west of the source. Subsequent observations
in 1997 February failed to confirm this extension (\cite{cor97}).
Both Fender et al.\ (1997) and Corbel et al.\ (1997) report a flat or
inverted radio spectrum.

\begin{table}
\caption{Journal of MOST observations of GX\,339$-$4 and final calibrated 
843 MHz flux densities.
\label{tbl1}}
{
\begin{tabular}{lccclr} 
\hline \\[-3mm]
Sequence &Date & N$_{\rm samp}$ & TJD & \multicolumn{2}{c}{S$_{843}$} \\ 
\multicolumn{1}{l}{number} &
\multicolumn{1}{c}{ }&
\multicolumn{1}{c}{ } &
\multicolumn{1}{c}{(mid-obs.)}  & 
\multicolumn{2}{c}{(mJy)} \\
\hline \\[-3mm]
1659483\rlap{$^*$}  & 94/04/25 & 1796 & 49498.123 & 9.81 & \llap{$\pm$}0.70 \\
1659484      & 94/06/01 &  898 & 49504.976 & 6.04 & 0.99 \\
1659481      & 94/06/02 &  690 & 49506.003 & 7.87 & 0.97 \\
1659482      & 94/06/03 & 1795 & 49507.097 & 6.93 & 0.75 \\
1659485      & 94/10/04 & 1019 & 49629.867 & 7.08 & 1.36 \\
1659486      & 94/10/17 & 1101 & 49642.821 & 6.24 & 1.19 \\
16594810     & 95/04/30 & 1101 & 49837.281 & \llap{$<$}2.0$^{\rm a}$ & ---\\
16594811\rlap{$^*$} & 95/06/02 & 1151 & 49871.099 & 5.26 & 0.66 \\
16594815     & 95/08/25 & 1796 & 49954.983 & 4.22 & 0.68 \\
16594813     & 95/10/14 &  981 & 50004.734 & 3.11 & 0.65 \\
16594816     & 96/03/15 & 1796 & 50158.313 & 3.21 & 0.72 \\
16594819\rlap{$^*$} & 96/03/31 & 1796 & 50174.269 & 2.22 & 0.67 \\
16594820\rlap{$^*$} & 96/04/06 & 1796 & 50180.253 & 2.98 & 0.65 \\
16594821     & 96/04/19 & 1796 & 50193.217 & 2.47 & 0.71 \\
16594822     & 96/04/30 &  898 & 50204.187 & \llap{$<$}2.0$^{\rm a}$ & --- \\
16594823     & 96/05/08 & 1796 & 50212.166 & 2.00 & 0.70 \\
16594824     & 96/05/17 & 1796 & 50221.141 & 2.47 & 0.65 \\
16594825\rlap{$^*$} & 96/05/20 & 1795 & 50224.133 & 2.31 & 0.60 \\
16594826     & 96/05/24 & 1796 & 50228.122 & 2.28 & 0.66 \\ 
16594827     & 96/06/07 & 1617 & 50242.108 & \llap{$<$}1.9$^{\rm a}$ & ---  \\
16594828     & 96/06/13 & 1796 & 50248.067 & \llap{$<$}1.9$^{\rm a}$ & --- \\
16594829     & 96/06/23 & 1791 & 50258.040 & 1.76 & 0.59 \\
16594830\rlap{$^*$} & 96/07/13 & 1791 & 50277.985 & 6.47 & 0.58 \\
16594831     & 96/07/28 & 1796 & 50292.944 & 6.70 & 0.68 \\
16594832\rlap{$^*$} & 96/08/06 & 1796 & 50301.920 & 6.02 & 0.56 \\
16594833     & 96/08/24 & 1619 & 50319.896 & 3.46 & 0.66 \\
16594834\rlap{$^*$} & 96/09/15 & 1796 & 50341.810 & 3.47 & 0.65 \\
16594836\rlap{$^*$} & 97/02/04 & 1795 & 50483.423 & 7.04 & 0.66 \\
16594837\rlap{$^*$} & 97/02/11 & 1794 & 50490.404 & 6.25 & 0.65 \\
16594838\rlap{$^*$} & 97/02/18 & 1795 & 50497.385 & 6.14 & 0.71 \\
\hline
\end{tabular}
}
  \begin{list}{}{}
     \item[$^{\mathrm{a}}$]3$\sigma$ upper limit.  
     \item[$^*$]Co-added to form image in Fig.\ \ref{fig1}.
  \end{list}
\end{table}

\section{Observations and Data Reduction}
The field of {\gx} was imaged by MOST at 843 MHz at irregular
intervals from 1994 April to 1997 February.  The journal of
observations is given in Table 1.  In order to maximise sensitivity we
observed in non-multiplexed mode (\cite{rob91}), giving a field size
of 23{\arcmin}$\times$31{\arcmin} and a synthesized beam of
43{\arcsec}($\alpha$)$\times$57{\arcsec}($\delta$) at the declination
of {\gx}.  Most observations were full 12-hour syntheses (1796 24-s
data samples) but several shorter integrations were found to be
acceptable for estimating flux densities.  Editing of the raw data was
necessary to remove terrestrial interference.  For completeness we
list the final number of data samples (N$_{\rm samp}$) in Table 1.

A CLEANed and self-calibrated image was formed from each synthesis
observation using standard procedures (\cite{c&y95}).  Position and
flux density calibration was based on short scans of a number of
strong unresolved sources before and after each synthesis observation
of {\gx}.  Experience has shown that this procedure is correct to
$<$1$''$ in position and $\sim$5\% rms in flux density (\cite{hun91}).

\begin{figure}
  \resizebox{\hsize}{!}{\includegraphics{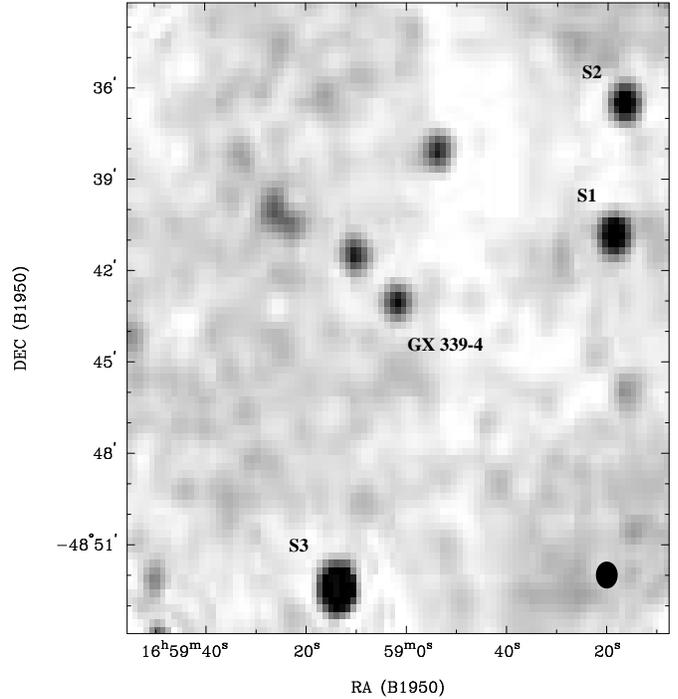}}
  \caption{ MOST 843 MHz image of the field of {\gx} obtained by
co-adding 11 separate images (see Table 1).  The reference sources
used for the final flux density calibration are marked, and the beam
is shown in the lower right corner. The weak negative north-south feature 
in the image is a grating response from an off-field soure.} \label{fig1}
\end{figure}

The AIPS task {\sc imfit} was
used to fit a gaussian to each image at the known position of {\gx}.
In a few cases where {\sc imfit} failed to converge (sequence numbers
16594810, 16594813, 16594822, 16594823, 16594828, 16594829) because of
a marginal detection, we used the task {\sc tvstat} to measure the
flux density. By integrating over two concentric regions centred on
{\gx} we were able to correct for non-zero or sloping backgrounds.

\begin{figure*}
  \resizebox{\hsize}{!}{\includegraphics{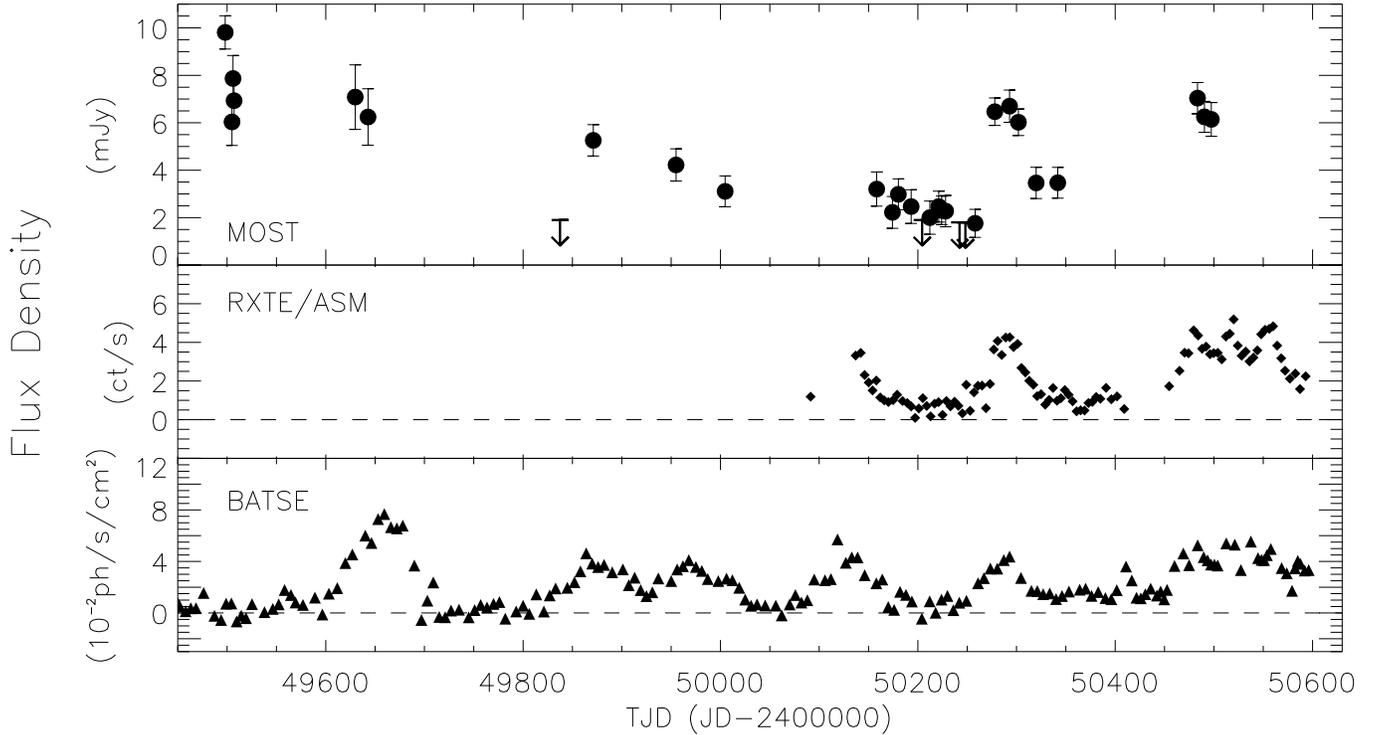}}
    \caption{
{\gx} light curves. The top panel shows the MOST light curve at
843 MHz, the middle panel the RXTE/ASM soft X-ray light curve
(2--10 keV) and the bottom panel depicts the hard X-ray light
curve as observed by BATSE (20--100 keV). The X-ray light curves
have been rebinned for clarity (see text).}
  \label{fig2}
\end{figure*}

In order to minimise systematic errors we then refined the
flux calibration by using {\sc imfit} to measure peak flux densities
for three reference sources S1, S2 and S3 (Fig. \ref{fig1}) in the
field of {\gx}.  All three were found to be unresolved, and to show
the same trends with epoch, so we adopted the summed flux density for
scaling the flux density of {\gx}, on the assumption that the MOST
calibration was correct on average over the monitoring period.  The
rms scatter of the correction factors was $\sim$5.2\%, consistent with
expectation.

The rms noise level in the vicinity of {\gx} was estimated from the
statistics of regions $\sim$500 pixels to the south
and west of {\gx}.  We have adopted the mean rms noise level (in
mJy/beam) as the flux density error for each epoch, and for setting a
limit for the non-detections.

The final corrected flux densities and adopted errors are tabulated in
Table \ref{tbl1}, while Table \ref{tbl2} gives the {\sc imfit}
positions and flux densities of the three reference sources.  The
truncated Julian date (TJD) is defined here as ${\rm TJD} = {\rm
JD} - 2400000$ and specified at mid-observation.  Note the fall in rms
noise between 1994 and 1995, resulting from the installation of
new pre-amplifiers as part of the hardware upgrade of MOST.

Figure \ref{fig1} shows a deep image of the field of {\gx} obtained by
co-adding selected images from the full dataset (see Table
\ref{tbl1}).  The background noise level of 0.3 mJy/beam rms 
includes contributions from weak sources, low level Galactic emission
and telescope artifacts.  There is no evidence for any prominent
extended radio emission which might be associated with {\gx}, in
contrast to a similar montage of the field around \object{{\rm
GRO\,J}1655$-$40}, comprising images taken before, during and after
the weak May-June 1996 outburst (\cite{hwcw}).

\begin{table}
\caption{Positions and mean flux densities
for the reference sources from the co-added image in Fig. 1.}
\label{tbl2}
\begin{tabular}{cccc} 
\hline
 & $\alpha$ (B1950.0) & $\delta$ (B1950.0) & $S_{843}$ \\
  & (h\hspace{1ex} m\hspace{1.5ex} s\hspace{2.8em} s ) & 
( {\degr}\hspace{1.5ex} {\arcmin}\hspace{1.5ex} {\arcsec}\hspace{2em} {\arcsec}) 
 &  (mJy) \\
\hline
S1 & 16 58 18.62 $\pm$ 0.14 & $-$48 40 50.6 $\pm$ 0.9 & 9.1 $\pm$ 0.3 \\
S2 & 16 58 16.64 $\pm$ 0.18 & $-$48 36 30.6 $\pm$ 1.3 & 8.0 $\pm$ 0.4 \\
S3 & 16 59 13.64 $\pm$ 0.09 & $-$48 52 26.5 $\pm$ 0.6 & 28.5 $\pm$ 0.6 \\
\hline\\[-8mm]
\end{tabular}
\end{table}

\section{Comparison with X-rays}
A relationship between hard X-ray and radio behaviour has been
established, though not well understood, for the two well-known
superluminal radio-jet X-ray binaries GRS\,1915+105 and
GRO\,J1655$-$40, and between soft (1--6 keV) X-ray and radio emission
for Cyg X-3, also known to exhibit milliarcsecond radio jets
(\cite{spencer86}).  In GRO\,J1655$-$40, the radio out\-bursts
appeared to follow flaring epi\-sodes in the hard (20--400 keV) X-ray
emission with a delay that varied from a few days to two weeks (Harmon
et al.\ 1995).  On the other hand, \object{GRS\,1915+105} exhibits
correlated hard (20--100 keV) X-ray and radio emission
(\cite{foster96}), with a clear but complex association between the
soft (2--10 keV) X-ray and radio (15 GHz) emission (\cite{poo97}).
Watanabe et al.\ (1994) report that radio outbursts from \object{Cyg
X-3} occur only when it is in the X-ray high state, i.e. the mean
soft X-ray flux is greater than usual.  McCollough et al.\ (1997) show
that during the quiescent radio state, the hard X-ray (20--100 keV)
flux of Cyg X-3 anticorrelates with the radio (15 GHz) flux, whereas
during radio flaring states the fluxes are sometimes correlated.  In
addition, a clear correlation between soft X-ray and 15 GHz radio 
emission has recently been revealed, e.g. Fender et al.
(1998).  Therefore, while there are obvious \mbox{\it connections\/}
between radio and X-ray emission from black-hole candidates, no
uniform pattern of behaviour has so far emerged.

Figure \ref{fig2} shows the final calibrated MOST light curve,
together with the RXTE/ASM\footnote{http://space.mit.edu/XTE/} and
BATSE\footnote{http://cossc.gsfc.nasa.gov/cossc/batse/hilev/GX339\_4/}
(C.\ Robinson, priv.\ comm.) light curves.  For clarity the X-ray light
curves have been resampled into 4-day bins for XTE and 6-day bins for
BATSE, using the IDL command {\sc rebin}.  Contrary to previous
results (e.g., \cite{sood97} and references therein), we find a
possible positive correlation between the X-ray and radio emission.

\begin{figure}
  \resizebox{\hsize}{!}{\includegraphics{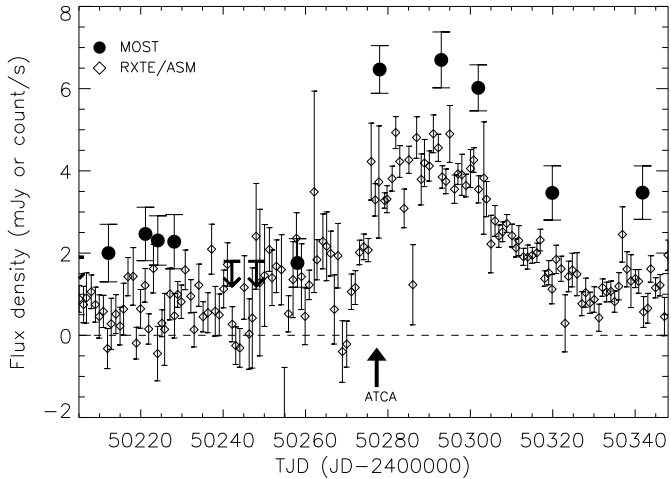}} 
\caption{
An overlaid plot of the {\gx} MOST and RXTE/ASM (2--10 keV) light
curves on an expanded scale.  The X-ray data points represent one-day
averages (1 ct/s $\simeq$ 13 mCrab).  The arrow at TJD $\sim$ 50277
marks the epoch when Fender et al.\ (1997) reported a possible radio
jet-like feature from ATCA observations.}
\label{fig3}
\end{figure}

Figure \ref{fig3} is an overlaid plot of the radio and soft X-ray
light curves covering the best sampled overlap region (1996 May--Sep).
Unfortunately, there was no radio coverage during the rapid rise in
the soft X-ray flux (beginning TJD $\sim$ 50270), so we cannot pinpoint
the exact time of the increase in the radio flux density.  However, it
is interesting to note that the radio light curve tracks the X-ray
curve quite closely, both before the small flaring episode and during
the maximum and decline.

\begin{figure}
  \resizebox{\hsize}{!}{\includegraphics{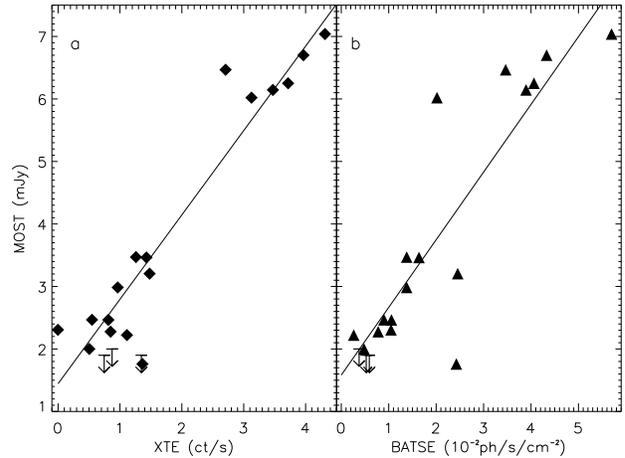}} \hfill
\caption{ 
Flux--flux plots of {\it (a)} MOST and RXTE, and {\it (b)} MOST and
BATSE data over the interval TJD = 50150--50500. In each case the
straight line is a least squares fit to the data points.}
\label{fig4}
\end{figure}

Figure \ref{fig4} shows flux--flux plots of the MOST and X-ray light
curves spanning the time interval from 1996 March to 1997 February.
The X-ray datasets were averaged or interpolated to the mean epochs of
the MOST observations, and a simple linear fit was applied.  The
resulting $\tilde{\chi}^2$ goodness-of-fit (15 d.o.f.) was 0.41 for
the MOST/RXTE fit and 0.99 for MOST/BATSE.  The Spearman rank
correlation ($\rho$) was computed for the two radio--X-ray datasets
using IDL's {\sc r\_correlate}.  For the MOST and RXTE data, $\rho$ is
0.82 and the two-sided significance of the deviation from zero is
$5.4\times 10^{-5}$.  The corresponding values for the MOST and BATSE
data are 0.82 and 4.9$\times 10^{-5}$.  These results indicate that
radio correlates well with both hard and soft X-ray intensities over a
time interval of nearly 1 year. In addition, given the limitations of
the radio sampling, there appears to be no evidence for a radio--X-ray
time delay for the event near TJD 50290.  On the other hand, it is
obvious that the first four MOST points (TJD $\sim$ 49500) do {\it not\/}
correlate with BATSE so there are clearly other factors affecting the
radio output at that time.  One possibility is we are witnessing the
decay of a much bigger radio event that may have been associated with
an earlier BATSE outburst at TJD $\sim$ 49400.

The fact that the ATCA 3.5\,cm observation (\cite{fen97}), at TJD
50276.79, was made just after the rapid rise in RXTE flux (Fig.\
\ref{fig3}) encourages speculation as to the possible origin of the
reported radio `jet'.  On the one hand it may be a real feature linked
directly to the X-ray and radio increase, while on the other it may be
an artifact introduced by phase errors in the telescope or changes in
source flux density.  The latter effect has indeed been seen in a MOST
observation obtained on 1997 July 22 in which an apparent elongation
of the image was clearly due to a variation in the flux density
of {\gx} within the 12-hour
integration period.

During the 1998 January high state transition observed by RXTE,
beginning at TJD $\sim$ 50820, both the MOST and BATSE fluxes fell to
zero (R.\ Fender, priv.\ comm.). This is very similar to the behaviour
observed in Cyg X-1, where transitions to the X-ray high (soft) state
are accompanied by a corresponding decrease in the radio flux density
(\cite{tan72}; \cite{bm76}), presumably related to changes in the
state of the accretion disk.  Furthermore, in the low state the X-ray
and radio fluxes of Cyg X-1 are roughly correlated (Pooley et al. 1998,
in prep.), as they appear to be in {\gx}.

\begin{acknowledgements}

We thank Barbara Piestrzynski for assistance with the MOST data
reduction, Craig Robinson for kindly providing pre-release BATSE data, and 
Rob Fender for valuable discussions during the drafting of the paper.
MOST is operated by the University of Sydney and funded by grants from
the Australian Research Council.  DH was funded by a grant from the
Vilho, Yrj\"o \& Kalle V\"ais\"al\"a Rahasto, and thanks the
Astrophysics Department of the University of Sydney for financial
assistance and hospitality during her visit.

\end{acknowledgements}


\end{document}